\documentclass[twocolumn]{aastex631}
\usepackage{amsmath}
\usepackage{bbm}

\newcommand{\unit}[1]{\,\,\mathrm{#1}}

\newcommand{\Msun}{M_\odot}
\newcommand{\SgrA}{Sgr A$^*$}
\newcommand{\ind}{\mathbbm{1}}

\begin{document}

\title{Detecting gravitational-wave bursts from black hole binaries in the Galactic Center with LISA}

\correspondingauthor{Alan M. Knee}
\email{aknee@phas.ubc.ca}

\newcommand{\ubc}{Department of Physics \& Astronomy, University of British Columbia, Vancouver, BC V6T 1Z1, Canada}
\newcommand{\ucla}{Department of Physics \& Astronomy, University of California, Los Angeles, CA 90095, USA}
\newcommand{\mani}{Mani L.~Bhaumik Institute for Theoretical Physics, Department of Physics \& Astronomy, UCLA, Los Angeles, CA 90095}
\newcommand{\camb}{Department of Applied Mathematics and Theoretical Physics, Cambridge CB3 0WA, United Kingdom}
\newcommand{\kavli}{Kavli Institute for Cosmology Cambridge, Madingley Road Cambridge CB3 0HA, United Kingdom}
\newcommand{\mon}{School of Physics and Astronomy, Monash University, VIC 3800, Australia}
\newcommand{\ozg}{OzGrav: Australian Research Council Centre of Excellence for Gravitational Wave Discovery, Clayton, VIC 3800, Australia}

\author[0000-0003-0703-947X]{Alan M. Knee}
\affiliation{\ubc}

\author[0000-0003-0316-1355]{Jess McIver}
\affiliation{\ubc}

\author[0000-0002-9802-9279]{Smadar Naoz}
\affiliation{\ucla}
\affiliation{\mani}

\author[0000-0002-4181-8090]{Isobel M. Romero-Shaw}
\affiliation{\camb}
\affiliation{\kavli}

\author[0000-0003-0992-0033]{Bao-Minh Hoang}
\affiliation{\ucla}
\affiliation{\mani}

\author[0000-0001-7113-723X]{Evgeni Grishin}
\affiliation{\mon}
\affiliation{\ozg}



\begin{abstract}
Stellar-mass black hole binaries (BHBs) in galactic nuclei are gravitationally perturbed by the central supermassive black hole (SMBH) of the host galaxy, potentially inducing strong eccentricity oscillations through the eccentric Kozai-Lidov (EKL) mechanism. These highly eccentric binaries emit a train of gravitational-wave (GW) bursts detectable by the Laser Interferometer Space Antenna (LISA)---a planned space-based GW detector---with signal-to-noise ratios (SNRs) up to ${\sim}100$ per burst. In this work, we study the GW signature of BHBs orbiting our galaxy's SMBH, Sgr A$^*$, which are consequently driven to very high eccentricities. We demonstrate that an unmodeled approach using a wavelet decomposition of the data effectively yields the time-frequency properties of each burst, provided that the GW frequency peaks between $10^{-3}\,\,\mathrm{Hz}$--$10^{-1}\,\,\mathrm{Hz}$. The wavelet parameters may be used to infer the eccentricity of the binary, measuring $\log_{10}(1-e)$ within an error of $20\%$. Our proposed search method can thus constrain the parameter space to be sampled by complementary Bayesian inference methods, which use waveform templates or orthogonal wavelets to reconstruct and subtract the signal from LISA data.
\end{abstract}



\section{Introduction} \label{sec:intro}

The Laser Interferometer Space Antenna (LISA) is a planned space-borne gravitational-wave (GW) observatory \citep{2017arXiv170200786A, Colpi:2024xhw}, currently set to launch in the late 2030s. With sensitivity in the $10^{-4}\unit{Hz}$ to $10^{-1}\unit{Hz}$ frequency range, LISA will unlock the source-rich mHz band of the GW spectrum, potentially detecting GWs from stellar-mass black hole binary (BHB) inspirals, massive BHB mergers, extreme mass-ratio inspirals, and millions of ultra-compact galactic binaries composed of white dwarfs and neutron stars \citep{2023LRR....26....2A}.
The ability of LISA to observe compact binaries long before they merge will present unique opportunities to study the dynamics of hierarchical multi-body systems \citep{2023LRR....26....2A}, where the influence of the perturbing body can have a measurable effect on the gravitational waveform, thus revealing key insights into compact binary formation channels in dense stellar environments \citep{1993Natur.364..423S, 2000ApJ...528L..17P, 2003ASPC..296...85M, 2016PhRvD..93h4029R, 2020A&A...640L..20B, 2021NatAs...5..749G, Bartos:2016dgn, Tagawa:2019osr}.

\begin{figure}
    \epsscale{1.15}
    \plotone{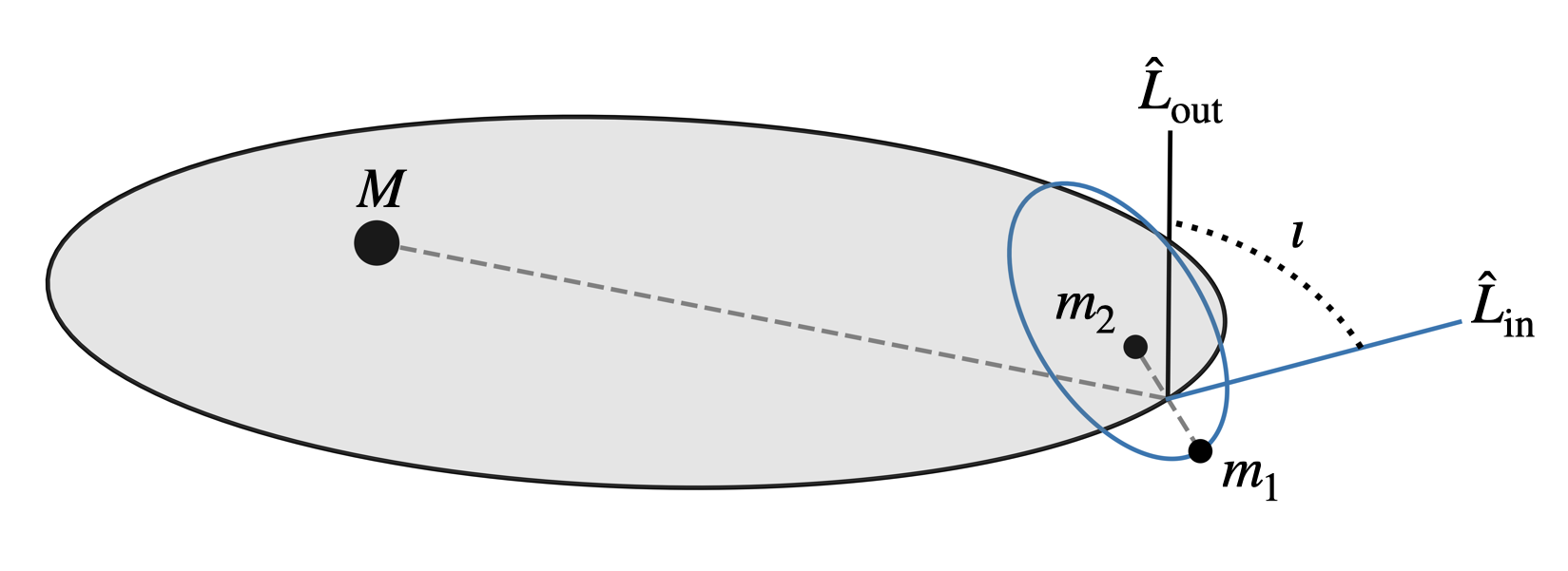}
    \caption{Schematic diagram of a hierarchical triple system. The inner binary (blue ellipse), consisting of two masses $m_1$ and $m_2$, orbits around a tertiary mass $M$ on a wide outer orbit (grey ellipse). The angle $\iota$ is the relative inclination between the inner and outer orbital planes. In this work, we focus on triples where the inner binary is a stellar-mass BHB and the tertiary body is a SMBH. \label{fig:diagram}}
\end{figure}

Galactic nuclei are expected to contain an abundance of stellar-mass black holes (BHs), many of which may exist in binaries. Two-body relaxation naturally causes heavier masses to migrate inwards, resulting in a dense compact object core where BHBs are readily assembled and hardened through repeated BH-BH encounters \citep{2006ApJ...649...91F, 2009MNRAS.395.2127O, 2016ApJ...824L..12O, 2016ApJ...831..187A, 2018PhRvD..98l3005R, 2018PhRvL.120o1101R, 2018PhRvD..97j3014S}. The Milky Way is no exception, potentially hosting ${\sim}2\times 10^4$ BHs within the inner parsec of the Galactic Center \citep{1993ApJ...408..496M, 2000ApJ...545..847M, 2006ApJ...649...91F, 2009MNRAS.395.2127O, 2018ApJ...853L..24N, 2022ApJ...929L..22R}. If the galaxy hosts a central supermassive black hole (SMBH), as is believed to be true for most large galaxies \citep{2013ARA&A..51..511K} including our own \citep{2005ApJ...620..744G, 2009ApJ...692.1075G}, stellar-mass BHBs may become bound to the SMBH on a relatively wide outer orbit, forming a hierarchical triple \citep{Antonini:2012ad, 2018ApJ...856..140H, 2018MNRAS.481.4907G, 2020ApJ...904..113R}. We illustrate this special orbital configuration in Fig.~\ref{fig:diagram}. Gravitational torques exerted by the SMBH can subsequently induce secular eccentricity and inclination oscillations through the eccentric Kozai-Lidov (EKL) mechanism, driving highly inclined BHBs towards eccentricities of nearly unity \citep{1962AJ.....67..591K, 1962P&SS....9..719L, 2016ARA&A..54..441N} and resulting in a GW source relevant for LISA. These novel sources possess a unique GW signature, as their extreme eccentricity concentrates the GWs into a train of bursts occurring once every periastron passage \citep{2009MNRAS.395.2127O, 2011ApJ...729L..23G, PhysRevD.85.123005, Xuan:2023azh}. Changes in the morphology and timing of the bursts reflect the secular evolution of the BHB due to the SMBH, allowing one to infer the orbital parameters of both the binary and its perturber \citep{2023PhRvD.107l2001R}. Eccentricity notably enhances GW emission and accelerates the inspiral timescale \citep{PhysRev.131.435}, possibly contributing to the compact binary merger rate observed by ground-based GW detectors \citep{2018ApJ...856..140H, 2019ApJ...878...58S, 2020PhRvD.102f4033L, 2020ApJ...903...67M, 2023PhRvX..13a1048A, 2019MNRAS.488...47F}.

Previous works have established that SMBH-induced BHB eccentricity oscillations are detectable by LISA out to a few Mpc \citep[e.g.,][]{ 2018ApJ...864..134R,2019ApJ...875L..31H, 2021ApJ...917...76W}\footnote{This effect can also take place in a wide range of masses and separations, see e.g., \citet{2019arXiv190702283R,2020MNRAS.495..536E,2020ApJ...901..125D}.}, provided the oscillation timescale is comparable to the mission lifetime. The loudest BHBs are expected to be orbiting the $4\times 10^6\,\,\Msun$ SMBH at the Galactic Center (\SgrA), with an estimated few tens of detectable sources in the LISA band \citep{2021ApJ...917...76W, Xuan:2023azh}. Due to their close proximity and heavy masses, these binaries can accumulate signal-to-noise ratios (SNRs) between $10^2$--$10^4$ over the course of a $4\unit{yr}$ LISA mission \citep{2019ApJ...875L..31H, 2021ApJ...917...76W}, necessitating efforts to model and subtract the bursts from LISA data \citep{2020PhRvD.101l3021L, 2023PhRvD.107f3004L}. In this Letter, we study the GW signature of BHBs orbiting \SgrA, prioritizing the initial task of {\it detection}. We calculate the GW signal in the time domain, including the full instrument response of LISA, by incorporating simulated background noise to emulate a detection scenario as realistically as possible. As a proof-of-concept, we explore the use of a wavelet decomposition, based on the multi-resolution $Q$-transform \citep{2004CQGra..21S1809C, chatterjithesis}, to resolve bursts from eccentric BHBs in LISA data and constrain their time-frequency properties. Additionally, we show that this time-frequency information provides a means to infer the orbital period and eccentricity of the BHB.


\section{Simulated LISA data} \label{sec:data}

\begin{figure*}
    \epsscale{1.15}
    \plotone{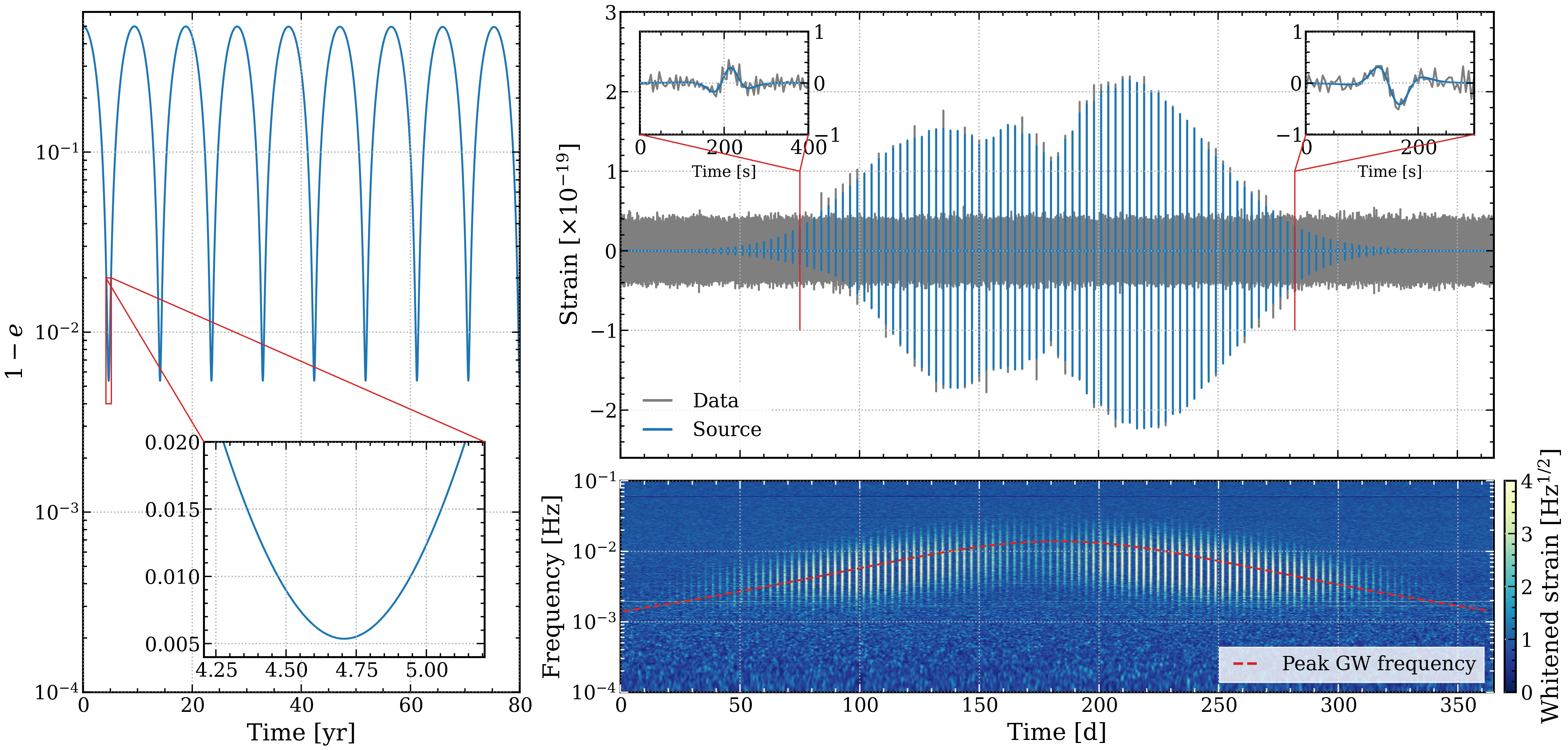}
    \caption{Simulated LISA observations of a singular BHB source with component masses $m_1=30\,\Msun$ and $m_2=20\,\Msun$, orbiting around a $M_\mathrm{SMBH}=4\times10^6\,\Msun$ SMBH. The initial parameters of the inner (outer) orbits are: semimajor axis $a=0.15\unit{au}$ ($a_\mathrm{out}=100\unit{au}$), eccentricity $e=0.5$ ($e_\mathrm{out}=0.01$), argument of periastron $\omega=0^\circ$ ($\omega_\mathrm{out}=0^\circ$), and inclination $\iota=85^\circ$ between the inner and outer orbits. The source is placed at a distance of $D=8\unit{kpc}$ with ecliptic latitude $\beta=-5.608^\circ$ and longitude $\lambda=266.852^\circ$, corresponding to the Galactic Center. The left panel shows the eccentricity evolution of the BHB over time. The inset focuses on a $1\unit{yr}$ interval centered on the time of maximum eccentricity ($e_\mathrm{max}\approx 0.9946$). The upper right panel shows the time-domain strain measured by LISA (in terms of the TDI-$A$ channel) during this same $1\unit{yr}$ period, where the signal of interest is shown in blue, and the total strain is in grey. The noise background is obtained from \texttt{LDC-2b} \citep{2022arXiv220412142B}. The lower right panel is a spectrogram of the whitened strain, with the peak GW frequency (Eq.~\ref{eq:fpeak}) represented by the dashed red curve. \label{fig:source}}
\end{figure*}

We begin by examining the GW signal originating from a stellar-mass BHB orbiting around a \SgrA-like SMBH, as observed by LISA. In order to generate simulations of the expected signal, we calculate the time-domain GW polarizations $h_{+,\times}$ for the BHB using a leading-order adiabatic approximation described in \citet{Barack:2003fp}, which expresses the waveform as a sum over harmonics of the orbital frequency:
\begin{equation} \label{eq:hsum}
    h_{+,\times}(t,\boldsymbol{\theta}) = \frac{G^2}{c^4D}\frac{m_1m_2}{a}\sum_n h_{+,\times}^{(n)}(t,\boldsymbol{\theta})\,,
\end{equation}
where $m_1$ and $m_2$ are the component masses, $D$ is the distance from the detector, and $\boldsymbol{\theta}$ denotes the standard (time-varying) Keplerian orbital elements. The mathematical definitions of the waveform harmonics $h_{+,\times}^{(n)}(t,\boldsymbol{\theta})$ are given in \citet{Barack:2003fp}. We evolve the system in time by numerically integrating the orbit-averaged equations of motion governing a hierarchical triple, including gravitational quadrupole and octupole-level contributions \citep{2011Natur.473..187N, 2013MNRAS.431.2155N}, general relativistic precession \citep{Naoz:2012bx}, and gravitational radiation \citep{PhysRev.136.B1224}. The explicit dynamical equations can be found in \citet{2016ARA&A..54..441N}. We then evaluate Eq.~\ref{eq:hsum} by sampling the time-evolved orbital elements. We ensure stability by imposing that the BHB cannot exceed its Hill radius \citep{2014ApJ...795..102N, 2018ApJ...856..140H, 2017MNRAS.466..276G, 2022PASA...39...62T}, as well as to avoid the breakdown of the secular approximation \citep{2016MNRAS.458.3060L, 2018MNRAS.481.4907G}. Further, the binaries we consider have hardened enough such that we expect them to resist disruption by the local stellar environment (see e.g., \citet{2018ApJ...856..140H,2020ApJ...904..113R} for discussion on the survivability of wide binaries in galactic nuclei).

In LISA, the output data streams are the time-delay interferometry (TDI) channels, e.g., the noise-orthogonal $\{A,E,T\}$ variables. These channels are synthesized by combining time-shifted one-way phase measurements recorded by each of LISA's six laser links \citep{Armstrong_1999, 2021LRR....24....1T}, and are designed to suppress the otherwise dominant laser noise. We use the instrument model implemented in \texttt{LISA GW Response} \citep{bayle_2022_6423436} to evaluate the interferometric phase shifts for each laser link given an incident GW signal. We use \texttt{PyTDI} \citep{staab_2022_6351737} to convert these measurements into the various TDI channels. Finally, we inject the TDI signal into mock LISA data from the {\it Spritz} edition of the LISA Data Challenges \citep[\texttt{LDC-2b},][]{2022arXiv220412142B}. For simplicity, we use a ``cleaned'' version of the $1\unit{yr}$ dataset, which is free of artifacts such as noise glitches and data gaps but retains various instrumental and environmental background noises.

Fig.~\ref{fig:source} shows $1\unit{yr}$ of simulated LISA observations for a representative $30\,\Msun+20\,\Msun$ BHB perturbed by a \SgrA-like SMBH and undergoing eccentricity oscillations. The signal is characterized by a train of repeating, short-duration pulses lasting a few minutes each, which map onto the frequency domain as broadband bursts of excess power. The burst cadence is equal to the orbital period of the BHB, which is about $3\unit{days}$ for the example shown in Fig.~\ref{fig:source}. The majority of the GW power is radiated near periastron, causing the spectrum of each to burst to peak at a frequency of approximately \citep{2009MNRAS.395.2127O}
\begin{equation}\label{eq:fpeak}
    f_\mathrm{p}(a, e) \approx \frac{(1+e)^{1/2}}{(1-e)^{3/2}}f_\mathrm{orb}(a)\,,
\end{equation}
where $(a,e)$ are the BHB semimajor axis and eccentricity at the time the burst was emitted, and $f_\mathrm{orb}(a)=(2\pi)^{-1}\sqrt{G(m_1+m_2)/a^3}$ is the orbital frequency. The burst amplitude and frequency track the oscillating eccentricity, as shown in the lower right panel of Fig.~\ref{fig:source}. The bursts also experience complicated amplitude modulations attributed to both the source's evolution and the motion of LISA, the latter of which depends on the source sky position and polarization. On account of the light travel time across LISA's heliocentric orbit (i.e.~Roemer delay), the burst arrival times measured by LISA deviate from the true orbital period by $\pm 500\unit{s}$, depending on the orbital phase of LISA.

Throughout Sec.~\ref{sec:data}--\ref{sec:param}, we assume the outer orbit of the triple is always coplanar with the sky (i.e. face-on/off inclination). For an inclined outer orbit, the GW signal will be Doppler shifted due to the projected motion of the inner binary along the line-of-sight as it orbits around the SMBH \citep{PhysRevD.109.064086}. This effect, known as the Roemer delay, will cause the time between bursts to oscillate about the orbital period, potentially biasing the estimate of the orbital period if the time delay represents a significant fraction of the inner orbital period. By assuming a coplanar outer orbit with the sky, we can ignore this Doppler shifting, which simplifies the data analysis problem. We provide a more detailed discussion of how the Roemer delay impacts our analysis in Appendix \ref{app}. We do not consider relativistic corrections from the motion of the BHB \citep{Robson:2018svj, Xuan:2022qkw} or the gravitational potential of the SMBH \citep{Kuntz:2022juv, Sberna:2022qbn}, opting to focus on a simple scenario in this proof-of-principle study and leave further generalization to future work.

\section{Unmodeled burst detection} \label{sec:search}

Motivated by their transient and burst-like nature, we adapt the $Q$-transform \citep{10.1121/1.400476} to detect highly eccentric BHBs in LISA data. Analysis pipelines based on the $Q$-transform search for arbitrary GW transients by filtering the data against windowed sinusoids ({\it wavelets}) in place of waveform templates \citep{2004CQGra..21S1809C, chatterjithesis}, and have been applied extensively to process data from ground-based GW detectors and characterize transient noise \citep{LIGO:2021ppb, Virgo:2022ysc}. We leverage the \texttt{GWpy} \citep{gwpy} implementation of the $Q$-transform, which is itself derived from earlier burst search pipelines \citep{2004CQGra..21S1809C, chatterjithesis, rollinsthesis, Robinet:2020lbf}. 

The $Q$-transform projects time series data onto overlapping time-frequency planes covered by an array of rectangular tiles, with each tile representing a wavelet. Each plane is parameterized by a quality factor, $Q \sim f/\Delta f$, which fixes the time-frequency aspect ratio of the individual tiles for that plane. The significance of any tile is quantified by its {\it energy}, given by \citep{2004CQGra..21S1809C, chatterjithesis}
\begin{equation}\label{eq:qtransform}
    |X(\tau,f,Q)|^2 = \bigg|\int_{-\infty}^\infty x(t)w(t-\tau,f,Q)e^{-2\pi i ft}\,\mathrm{d}t\bigg|^2\,,
\end{equation}
where $(\tau,f,Q)$ are the time, frequency, and $Q$ of the tile, $x(t)$ is the whitened strain data, and $w(t,f,Q)$ is a window function (specifically, a bisquare window). We normalize the energy such that, in white noise, $|X(\tau,f,Q)|^2$ is exponentially distributed with unit mean and variance. The tile density is tuned to guarantee a fractional energy loss ({\it mismatch}) between adjacent tiles no larger than a desired amount, with smaller mismatches resulting in denser time-frequency tilings. The analysis outputs all tiles with SNR exceeding a pre-determined threshold, called {\it triggers}, where we define the trigger SNR as
\begin{equation}\label{eq:snr}
    \hat{\rho} = \sqrt{|X(\tau,f,Q)|^2 - 1}\,.
\end{equation}
Because a single burst typically produces multiple triggers, we group significant triggers that are overlapping or adjacent in time into clusters, and report the maximum-SNR trigger for each cluster. In our implementation, the $Q$-transform is carried out in two stages: the first stage performs a ``quick pass'' of the data using a low-resolution $Q$-tiling at $15\%$ mismatch; after clustering triggers with SNR greater than $8$, we search $5\unit{hr}$ of data around each trigger with a high-resolution tiling at $1\%$ mismatch to further refine the trigger parameters.

Prior to calculating the $Q$-transform, we whiten the data by its power spectral density (PSD), which is estimated empirically via a Welch median \citep{welch} with $2^{19}\unit{s}$ Hann-windowed and $50\%$ overlapping segments. To minimize over-whitening, we calculate the PSD on a ``gated'' version of the data \citep{selfgating, Usman:2015kfa}. This gating is done by first dividing the data into short segments of duration $T_\mathrm{gate}=300\unit{s}$; if any of the strain values exceed some threshold, the data are multiplied by an inverse Tukey window, which is zero during the segment containing the excursion, and smoothly transitions from zero to one over a duration $T_\mathrm{gate}/4$ on either side. Through trial and error, we find that a gating threshold of $4.5$ times the root-mean-square strain (averaged over the full time series) is sufficient to remove the worst-offending bursts and prevent over-whitening.

\begin{figure*}
    \epsscale{1.15}
    \plotone{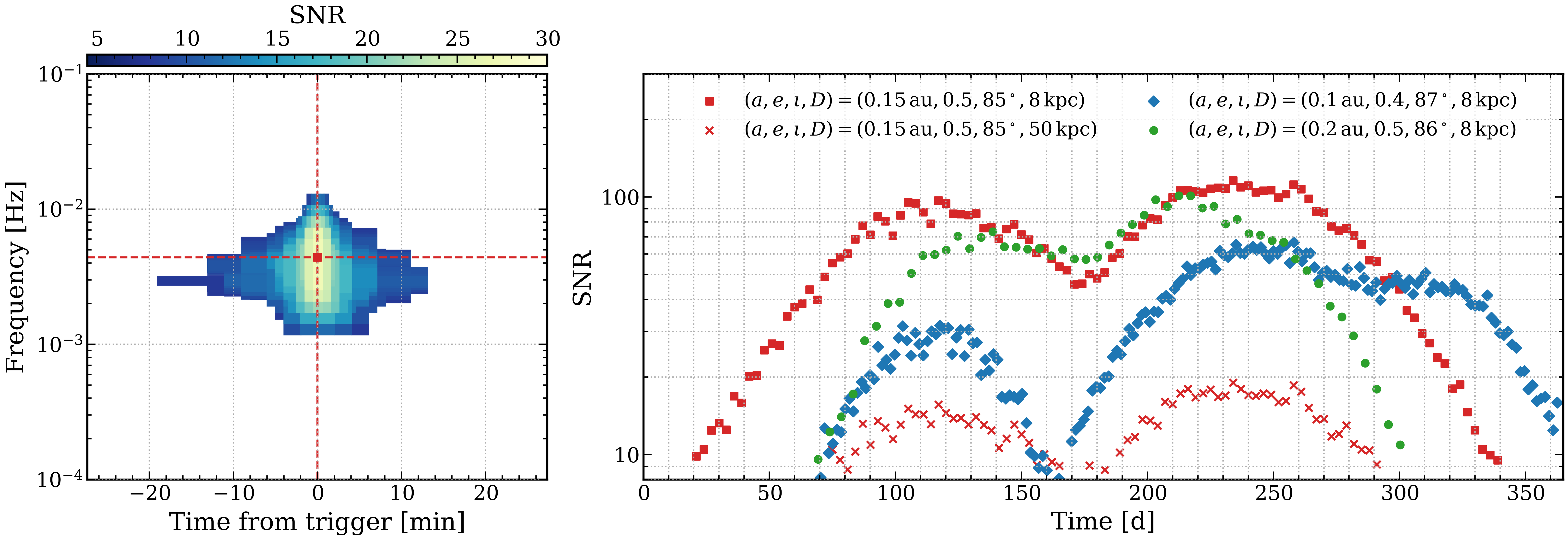}
    \caption{Time-frequency analysis of LISA data using the $Q$-transform. The data consists of simulated noise \citep[\texttt{LDC-2b},][]{2022arXiv220412142B} with added signals from three BHBs with masses $m_1=30\,\Msun$ and $m_2=20\,\Msun$ orbiting a $M_\mathrm{SMBH}=4\times 10^6\,\Msun$ SMBH at the Galactic Center. A fourth source located $50\unit{kpc}$ away is also shown. The initial BHB parameters are given in the legend. The initial outer orbit is specified by $a_\mathrm{out}=100\unit{au}$ and $e_\mathrm{out}=0.01$ for each source. The left panel is a visualization of the raw $Q$-transform output ($15\%$ mismatch, SNR threshold $8$) around a burst, where the time axis is given in minutes from the trigger time ($t_\mathrm{trig}\approx 54.02\unit{d}$). The SNR of each tile is given by its color. The square in the centre is the maximum-SNR trigger for this cluster. The right panel shows the time and SNR of all triggers ($1\%$ mismatch) with SNR above $8$, each one corresponding to a detected burst. \label{fig:triggers}}
\end{figure*}

Fig.~\ref{fig:triggers} shows all triggers with SNR above $8$ after analyzing simulated LISA data with the $Q$-transform. For simplicity, we consider just a single TDI channel\footnote{Our analysis could be improved by using additional TDI channels to perform glitch rejection, as true GWs will propagate through LISA differently from instrumental glitches \citep{Robson:2018jly}.}, TDI-$A$. We search over time-frequency planes with low $Q$-values\footnote{The lower bound of $Q=\sqrt{11}$ is an anti-aliasing condition.}, between $\sqrt{11}$ and $32$, specifically targeting bursts with short durations (maximizing timing accuracy) and high bandwidths. At $15\%$ ($1\%$) mismatch, this gives a total of $4$ ($14$) $Q$-planes. The searched frequency range is from $10^{-4}\unit{Hz}$ to $10^{-1}\unit{Hz}$, covering the nominal LISA observing band. The data contains three sources representing BHBs orbiting a SMBH at the Galactic Center, amidst background LISA noise from \texttt{LDC-2b} \citep{2022arXiv220412142B}. We also simulate a fourth source identical to the example from Fig.~\ref{fig:source}, but placed at a distance of $50\unit{kpc}$ (roughly the distance to the Large Magellanic Cloud) instead of $8\unit{kpc}$, showing how these more distant bursts are still individually resolvable at sufficiently large eccentricities. Three of the injected sources (red and green markers) achieve maximum eccentricity at the midpoint of the simulated observation period. We note that aligning the time of maximum eccentricity with the midpoint of the observation period is done merely to simulate an ``optimal'' detection scenario. Shifting the source in time will reduce the number of detected bursts. As an example, we include in Fig.~\ref{fig:triggers} another source (blue markers) which achieves maximum eccentricity at the end of the observation period, and so the source is only observed while its eccentricity is increasing. Longer mission lifetimes not only increase the chances that LISA will see one of these putative sources during a high-eccentricity state, but will also improve LISA's ability to study the source evolution over time and determine whether the dynamics are consistent with EKL-induced oscillations.

An important limitation of the $Q$-transform is that it is only sensitive to well-localized excess power, and thus cannot combine the power from multiple bursts far apart in time. Even if the summed SNR from all bursts is above our threshold, the source will not be detected if the SNR of {\it individual} bursts is consistently below the detection threshold. Our approach could thus be improved upon by stacking power from multiple bursts assuming some burst timing model \citep{2023PhRvD.107l2001R, Arredondo:2021rdt}, allowing one to filter against a train of wavelets instead of isolated wavelets.

\section{Parameter recovery} \label{sec:param}

Even though the $Q$-transform only provides generic time-frequency information about the bursts, it is possible to make rough inferences about some of the source's orbital parameters based on two key factors: the timing of the bursts, and the characteristic ``peak'' frequency where most of the excess power is concentrated. We outline here a simple method of estimating the orbital period and eccentricity evolution of BHBs using only the trigger time-frequency parameters.

\subsection{Orbital period} \label{sec:period}

\begin{figure*}
    \epsscale{1.15}
    \plotone{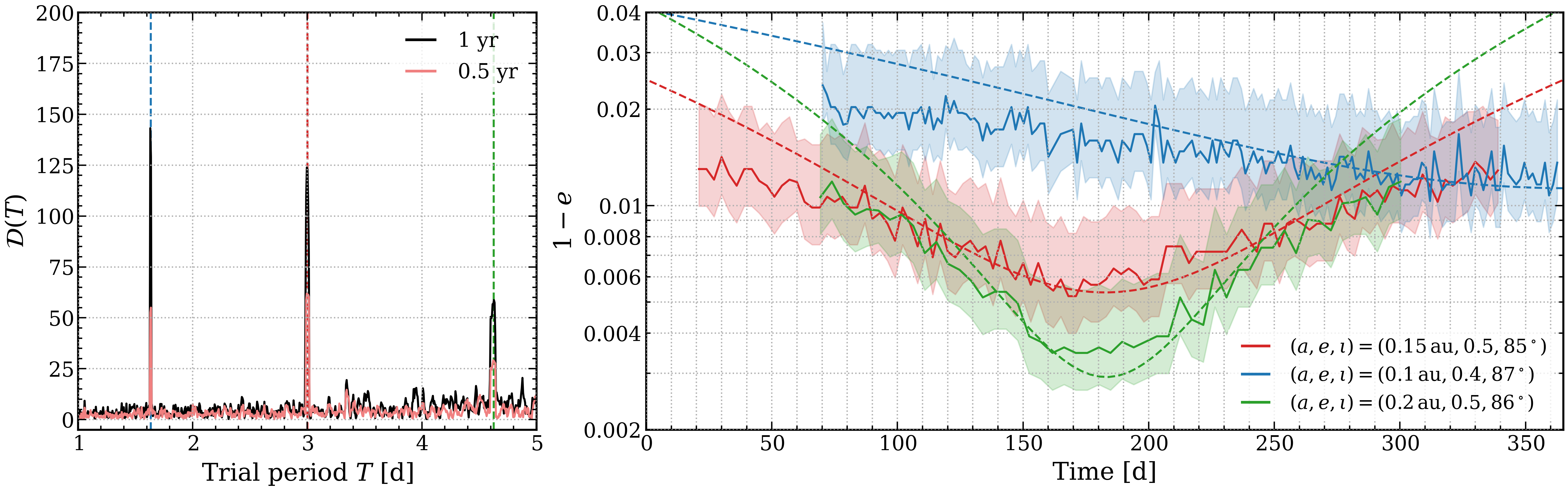}
    \caption{Recovered orbital period and eccentricity for three simulated BHBs with masses $m_1=30\,\Msun$ and $m_2=20\,\Msun$ orbiting a $M_\mathrm{SMBH}=4\times 10^6\,\Msun$ SMBH at the Galactic Center. The initial orbital parameters are repeated from Fig.~\ref{fig:triggers}. The left panel shows periodograms of simulated LISA triggers over $0.5$ (light red) and $1\unit{yr}$ (black) observing periods, calculated via Eq.~\ref{eq:dstat} with period step size $\Delta T=500\unit{s}$ and jitter $\epsilon=0.005$. The right panel shows the reconstructed eccentricity evolution of each BHB inferred from the $Q$-transform triggers. The solid lines are the eccentricity calculated from the central frequencies of each trigger, and the shaded region represents the uncertainty due to the bandwidth of each trigger. The true eccentricities are shown by the dashed curves. \label{fig:period_ecc}}
\end{figure*}

Bursts from highly eccentric binaries occur once per orbit at periastron, thus we can measure the orbital period by observing the time interval between consecutive burst triggers. However, the $Q$-transform is also likely to detect non-repeating events associated with, e.g., massive BHB mergers or detector glitches, obfuscating the eccentric BHB signal. In addition, LISA may detect bursts from more than one eccentric binary, introducing multiple periodicities into the trigger catalog. Numerous period detection schemes have been studied in the astronomical literature \citep{10.1093/mnras/stw848}. These methods typically involve the phase-folding of arrival time data over many trial periods and evaluating a test statistic, e.g.~the Rayleigh or $H$-test \citep{1989A&A...221..180D}, at each trial. Here, we apply a period-finding statistic from \citet{845217} to identify repeating bursts from individual binaries,
\begin{equation}\label{eq:dstat}
    \mathcal{D}(T) = \bigg|\sum_{k=2}^{N_\mathrm{t}}\sum_{l=1}^{k-1}\ind(|T-(t_k-t_l)|<\epsilon T)e^{2\pi i t_k/T}\bigg|\,,
\end{equation}
where $T$ is a trial orbital period, $0<\epsilon<1$ is a small jitter fraction, $t_i$ are the trigger times, $N_\mathrm{t}$ is the total number of triggers, and $\ind(\cdot)$ is the indicator function, which has a value of $1$ when its argument is true and $0$ otherwise. Eq.~\ref{eq:dstat} essentially iterates over all pairs of triggers $(t_k,t_l)$ and counts the number of pairs separated by an interval between $T(1\pm\epsilon)$. 

The $\mathcal{D}(T)$ statistic has a number of advantages that make it well-suited for our problem: the jitter fraction $\epsilon$ can be adjusted to admit a certain level of deviation from the expected period; the weighting by $e^{2\pi i t/T}$ suppresses higher harmonics of the fundamental period, allowing for easier identification of distinct sources; lastly, it is robust to one-off transients like glitches or mergers, assuming these occur randomly\footnote{Results from LISA Pathfinder showed that the waiting time between acceleration glitches was exponentially distributed \citep{2022arXiv220511938L}. However, it is not inconceivable that a full-size LISA could experience periodic glitching if the glitch source is periodic in nature. The efficacy of our proposed method will thus depend on how well such glitches are mitigated.}. The main downside to Eq.~\ref{eq:dstat} is that it checks {\it all} trigger pairs, and so the computational cost scales as $N_\mathrm{t}^2$. Also, if there are many missing bursts, the value of $\mathcal{D}(T)$ at the true period will be reduced.

Evaluating Eq.~\ref{eq:dstat} on the BHB triggers from Fig.~\ref{fig:triggers} over many trial periods gives a periodogram, with sharp peaks corresponding to the orbital periods of each binary, as shown in the left panel of Fig.~\ref{fig:period_ecc}. To test the robustness of this period search method in the presence of extraneous triggers, we artificially add $500$ triggers distributed in time with uniform probability between the start and end of the observing period. Even though the number of non-periodic triggers is comparable to the number of periodic triggers, each BHB period is clearly detected above the noise level. For comparison, we repeat our analysis with triggers from the first $0.5\unit{yr}$ of data, showing how the peak prominence rises linearly with the number of detected bursts. Note that we fix $\epsilon=0.005$, which limits the period deviation to $0.5\%$ of the folded trial period. Since we do not simulate the orbit of the inner binary around the SMBH, the only source of period deviation is due to the travel time across the orbit of LISA. Increasing $\epsilon$ allows Eq.~\ref{eq:dstat} to capture larger variations in the period, but at the expense of a higher noise level. Larger simulations are needed to flesh out the noise distribution of $\mathcal{D}(T)$ and determine appropriate thresholds for claiming evidence for periodicity.


\subsection{Eccentricity evolution} \label{sec:ecc}

We reconstruct the BHB eccentricity by leveraging the relation between its orbital frequency and peak GW frequency, as given in Eq.~\ref{eq:fpeak}. For the orbital frequency $f_\mathrm{orb}$ we substitute the value measured via Eq.~\ref{eq:dstat}, and for the peak GW frequency $f_\mathrm{p}$ we substitute the central frequency of the maximum-SNR trigger for each burst. A simple inversion of Eq.~\ref{eq:fpeak} yields an estimate of the eccentricity at the time each burst was emitted, which lets us examine the EKL-induced eccentricity evolution over time. 

We show reconstructed eccentricity tracks of simulated BHBs orbiting a SMBH at the Galactic Center in the right panel of Fig.~\ref{fig:period_ecc}. The eccentricity estimator roughly tracks the true eccentricity near the eccentricity peak, but is prone to overestimating at lower eccentricities, particularly at peak GW frequencies near $f_\mathrm{p}\sim 10^{-3}\unit{Hz}$. Further inspection shows that the peak frequency measured by LISA deviates from the true peak frequency at lower frequencies. This is because the bursts are effectively filtered by the TDI response functions of LISA \citep[e.g.,][]{Cornish:2002rt, Flauger:2020qyi}, which have a strong frequency dependence that shifts the detector-frame peak frequency to higher frequencies. The whitening filter follows a similar shape to the response function and compensates somewhat for this effect, but is not sufficient to shift the measured peak frequency back to the source-frame value for all frequencies. The effect of the LISA/TDI response functions on transient broadband signals should be studied in future work.

\begin{figure}
    \epsscale{1.15}
    \plotone{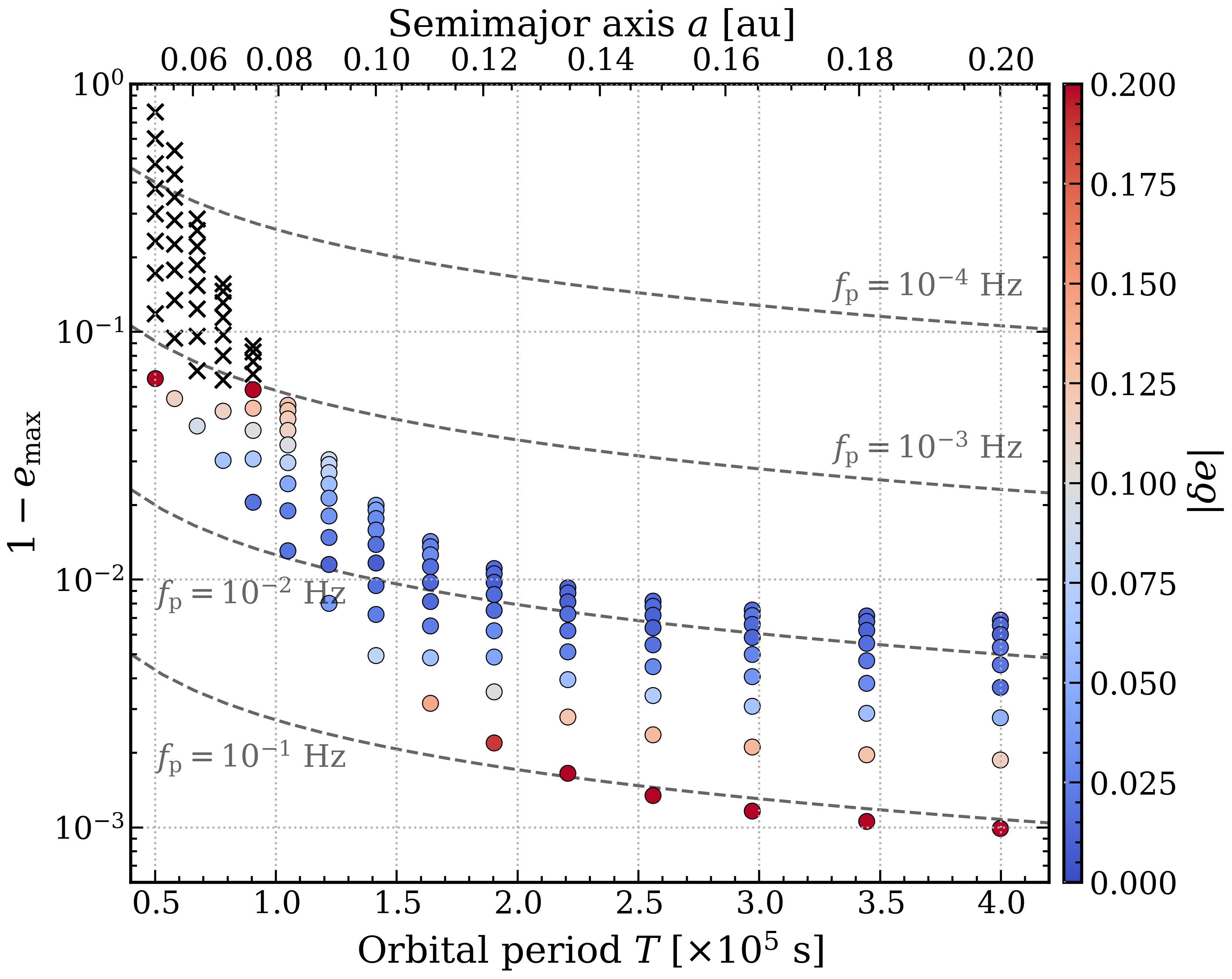}
    \caption{Eccentricity errors, $|\delta e|$ (Eq.~\ref{eq:err}), for simulated BHBs with masses $m_1=30\,\Msun$ and $m_2=20\,\Msun$ orbiting a $M_\mathrm{SMBH}=4\times 10^6\,\Msun$ at the Galactic Center. As before, the initial outer orbital parameters are $a_\mathrm{out}=100\unit{au}$, $e_\mathrm{out}=0.01$, and $\iota=85^\circ$ in all simulations. Each source is plotted at its maximum eccentricity, and $|\delta e|$ (marker colours) is averaged over a month of data centered on this time. For simulations with fewer than two detected bursts (black cross), we cannot measure the orbital period or eccentricity. The dashed grey contours are lines of constant peak GW frequency assuming a $30\,\Msun+20\,\Msun$ BHB. \label{fig:ecc_grid}}
\end{figure}

To better understand the accuracy of our eccentricity estimator over the parameter space, we generate $135$ galactic BHB simulations on a grid in initial semimajor axis $a\in[0.05\unit{au},0.2\unit{au}]$ and eccentricity $e\in[0.1,0.9]$, and recover the eccentricity of each source using our trigger-based method. The outer orbit remains the same for all simulations. Fig.~\ref{fig:ecc_grid} shows the relative eccentricity errors for our simulations in terms of
\begin{equation} \label{eq:err}
    |\delta e| = \bigg|1-\frac{\log_{10}(1-e_\mathrm{meas}(t))}{\log_{10}(1-e_\mathrm{true}(t))}\bigg|,
\end{equation}
where $e_\mathrm{meas}(t)$ and $e_\mathrm{true}(t)$ are respectively the measured and true eccentricity at trigger time $t$. Eq.~\ref{eq:err} is averaged over a month of data centered on the time of maximum eccentricity for each simulation. Overall, the relative error is less than $20\%$ across all simulations, and the error is lowest for peak frequencies near $f_\mathrm{p}\sim 10^{-2}\unit{Hz}$. At lower frequencies, the eccentricity becomes overestimated for the reasons discussed above. Moreover, the eccentricity is underestimated as the peak frequency approaches the Nyquist limit of $0.1\unit{Hz}$, where our frequency resolution quickly deteriorates on account of the logarithmic frequency scaling of the $Q$-transform. This low sample rate hinders our ability to characterize bursts from the most eccentric BHBs in our suite of simulations.

\section{Summary} \label{sec:disc}

A detection of binary eccentricity oscillations would offer conclusive evidence that the binary existed in a triple, providing key insight into the formation and evolution of stellar-mass compact binaries in dense cluster environments. In this work, we have studied the GW signature of stellar-mass BHBs in the Galactic Center that are driven to high eccentricities by a perturbing SMBH, and outlined techniques for detecting and identifying these sources in LISA data. BHBs in our own galaxy offer the best chances to detect EKL-driven dynamics, though we expect such systems to exist in other galaxies hosting SMBHs as well. We showed that an unmodeled time-frequency analysis can detect GW bursts emitted by highly eccentric BHBs in our galaxy, and characterize their time-frequency properties well enough to provide a rough estimate of the BHB orbital period and eccentricity evolution. This information can be passed to downstream analyses which use wavelets or waveform models to fit the signal, reducing the parameter space to be explored by these more computationally expensive techniques. Although we have focused on eccentric BHBs perturbed by a SMBH, our analysis techniques could be straightforwardly applied to generically eccentric compact binaries which do not undergo EKL oscillations. We worked with $1\unit{yr}$ of simulated LISA data for this study, but emphasize that longer observing times provide better opportunities to detect and study the long-term behaviour of EKL triples.

As this is a proof-of-concept exploration, there is much room to build upon and optimize various aspects of our methodology. Our burst search pipeline is based on the $Q$-transform, which uses bisquare-windowed sinusoids to maximize detection efficiency. However, this wavelet basis is overcomplete and cannot be used for direct signal reconstruction and subtraction, which will likely be a necessary step for the LISA global fit \citep{2020PhRvD.101l3021L, 2023PhRvD.107f3004L}. Future studies should explore alternative bases more suitable for burst subtraction, such as orthogonal Meyer wavelets, as done in the \texttt{cWB} pipeline \citep{Klimenko:2004qh, Klimenko:2008fu, Drago:2020kic}, or Morlet-Gabor wavelets with a manually enforced orthogonality condition, as done in \texttt{BayesWave} \citep{2015CQGra..32m5012C, 2021PhRvD.103d4006C}. Another limitation of our method is that we search for bursts in isolation and do not stack power coherently from multiple bursts. Our sensitivity could potentially be enhanced by filtering on templates which chain together many bursts, where the timing between bursts and their relative amplitudes represent fit parameters.

\begin{acknowledgments}

This research was enabled in part by support provided by the Digital Research Alliance of Canada (\url{alliance.can.ca}). A.M.K. and J.M. acknowledge funding support from the Natural Sciences and Engineering Research Council of Canada. A.M.K. acknowledges funding support from a Killam Doctoral Scholarship. J.M. acknowledges support from the Canada Research Chairs program. S.N. acknowledges the partial support from NASA ATP 80NSSC20K0505 and from NSF-AST 2206428 grant and thanks Howard and Astrid Preston for their generous support. I.M.R.-S. acknowledges support received from the Herchel Smith Postdoctoral Fellowship Fund. 

\end{acknowledgments}

%



\software{\texttt{GWpy} \citep{gwpy}, \texttt{SciPy} \citep{2020SciPy-NMeth}, \texttt{NumPy} \citep{harris2020array}, \texttt{Matplotlib} \citep{Hunter:2007}, \texttt{LISA GW Response} \citep{bayle_2022_6423436}, \texttt{PyTDI} \citep{staab_2022_6351737}.}



\appendix

\section{Effect of Roemer delay} \label{app}

In general, the GW signal from binaries in triples will be Doppler shifted due to the (time-varying) travel time across the outer orbit, i.e.~the Roemer delay. In the context of LISA, there are two contributions to this delay: the outer orbit of the hierarchical triple, and LISA's heliocentric orbit. The latter is small compared to the delay from the outer orbit, and so we focus only on the contribution from the outer orbit. We have thus far assumed the outer orbit to be coplanar with the sky, where there is no Roemer delay since the outer orbit is viewed face-on. In this section, we perform additional simulations in which this assumption is relaxed, and discuss the effectiveness of our period/eccentricity estimation method when the Roemer delay is present. 

Following \citet{Barack:2003fp}, we model the Roemer delay by shifting the signal phase according to
\begin{equation}
    \Phi(t) \rightarrow \Phi(t) + 2\pi f_\mathrm{orb}\frac{R_\mathrm{out}}{c}\sin\iota_\mathrm{out}\sin(\nu_\mathrm{out}+\omega_\mathrm{out})\,,
\end{equation}
where $f_\mathrm{orb}$ is the inner orbital frequency, $R_\mathrm{out}$ is the separation between the inner binary and SMBH, $\iota_\mathrm{out}$ is the inclination of the outer orbit, $\nu_\mathrm{out}$ is the true anomaly of the inner binary along the outer orbit, and $\omega_\mathrm{out}$ is the argument of pericenter of the outer orbit. For GW bursts, this Doppler shift manifests as a delay in the burst arrival time at the Solar System barycenter (SSB), equivalent to
\begin{equation}\label{eq:delay}
    t_\mathrm{a} = t_\mathrm{e} - \frac{R_\mathrm{out}}{c}\sin\iota_\mathrm{out}\sin(\nu_\mathrm{out}+\omega_\mathrm{out})\,,
\end{equation}
where $t_\mathrm{a}$ is the time of reception at the SSB, and $t_\mathrm{e}$ is the burst emission time. Note that when rotating the outer orbit to achieve certain inclinations, the same rotations are applied to the inner orbit, which incidentally affects the observed polarization of the GW bursts emitted by the inner binary.

\begin{figure}
    \epsscale{1.15}
    \plotone{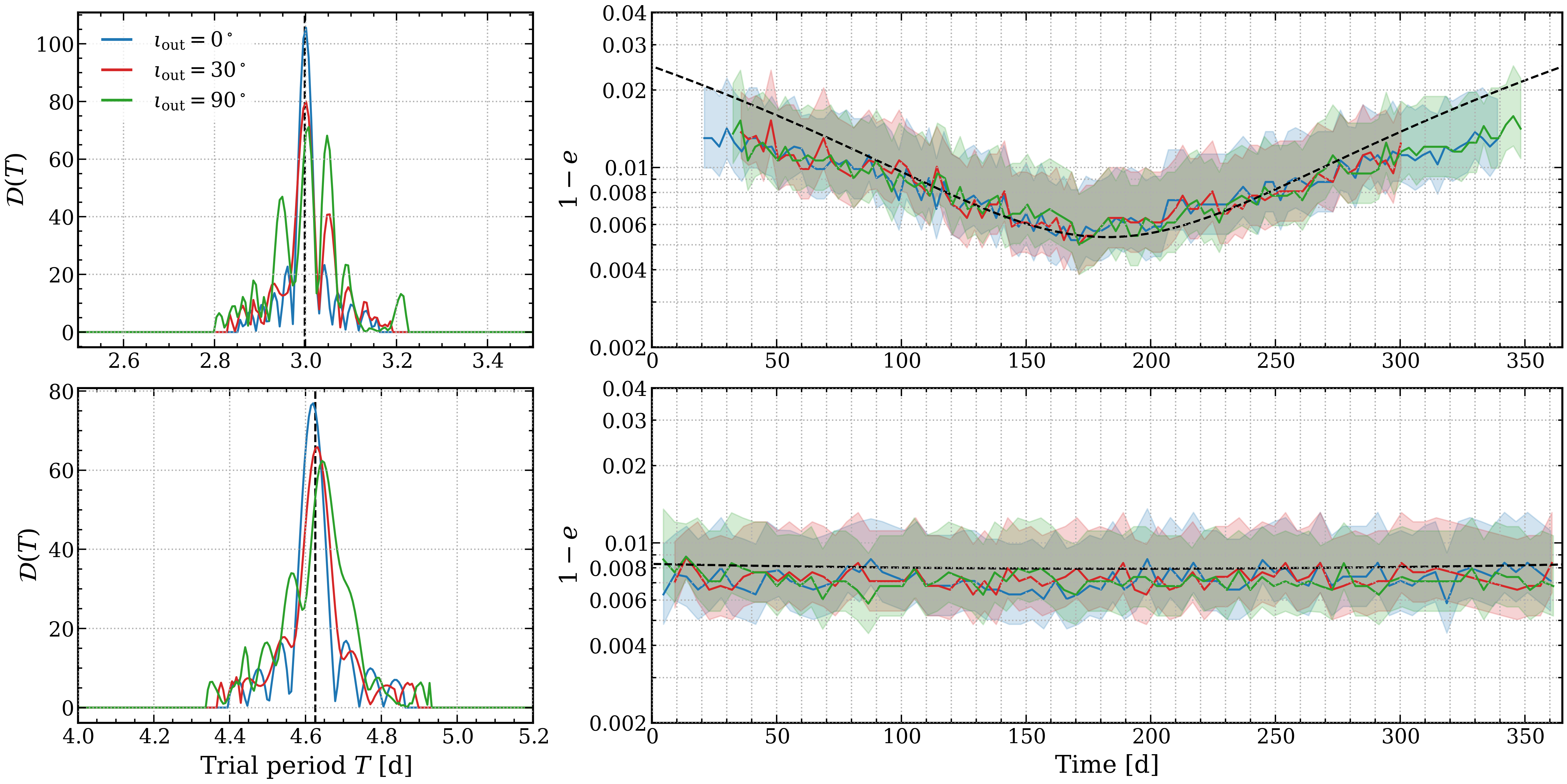}
    \caption{Period and eccentricity recovery for two simulated $30\,\Msun+20\,\Msun$ BHBs orbiting a \SgrA-like SMBH at the Galactic Center, assuming different outer orbit inclinations relative to the line-of-sight. The initial outer orbit for the system in the top (bottom) row is given by $a_\mathrm{out}=100\unit{au}$ ($a_\mathrm{out}=250\unit{au}$) and $e_\mathrm{out}=0.01$, and the initial inner orbit is given by $a=0.15\unit{au}$ ($a=0.2\unit{au}$), $e=0.5$ ($e=0.7$), and $\iota=85^\circ$ ($\iota=86^\circ$). The left panels show periodograms of the $Q$-transform triggers obtained for each source, where the dashed vertical lines are the inner orbital period. The right panels are the estimated eccentricities if the orbital period is inferred from the highest peak in the respective periodogram. The dashed black curves show the true eccentricity evolution. \label{fig:inc_grid}}
\end{figure}

The results of our simulations accounting for Roemer delay are summarized in Fig.~\ref{fig:inc_grid}. We select two example systems and inject the GW signal into LISA noise assuming three different choices of outer orbit inclinations: $\iota_\mathrm{out}=0^\circ$ (face-on), $30^\circ$, and $90^\circ$ (edge-on), where $90^\circ$ induces the maximum amount of delay. The left panels of Fig.~\ref{fig:inc_grid} are periodograms of the resulting $Q$-transform triggers. In calculating $\mathcal{D}(T)$ via Eq.~\ref{eq:dstat}, we set $\epsilon=0.05$ to accommodate the variation in the burst arrival times. For outer orbits with greater inclination relative to the line-of-sight, the Roemer delay introduces a multi-modal structure in the periodogram, due to the time interval between bursts varying sinusoidally about the inner orbital period according to Eq.~\ref{eq:delay}.

Our results show that the Roemer delay can potentially bias the estimated inner orbital period by a few percent. For fixed SMBH mass, the delay is larger for smaller outer orbits due to the faster orbital velocity of the outer orbit. It is also larger for inner binaries with longer orbital periods, since the phase shift accumulates between bursts. As shown by the right panels of Fig.~\ref{fig:inc_grid}, we find that the bias in the period is not large enough to impact the eccentricity measurement for our two example systems. However, the reduced height of the periodogram peaks due to the Roemer delay implies that if the set of triggers contains extraneous transients (e.g.~glitches), then the evidence for periodicity could be hidden. Perhaps the most robust method to deal with the Roemer delay is to fit a timing model to the trigger times which incorporates this time delay \citep{2023PhRvD.107l2001R}, thereby allowing one to recover the correct orbital period as well as estimate the parameters of the outer orbit, though we leave such endeavours to future work.


\bibliography{main}{}
\bibliographystyle{aasjournal}



\end{document}